\documentclass[11pt,twoside]{article}
\usepackage{asp2010}

\resetcounters

\begin{document}

\title{Helium abundances and the helium isotope anomaly of sdB stars}
\author{S. Geier$^1$, U. Heber$^1$, H. Edelmann$^1$, L. Morales-Rueda$^2$, D. Kilkenny$^3$,\\ D. O'Donoghue$^{4,5}$, T. R. Marsh$^6$, C. Copperwheat$^6$
\affil{$^1$Dr. Karl Remeis-Observatory \& ECAP, Astronomical Institute,
Friedrich-Alexander University Erlangen-Nuremberg, Sternwartstr. 7, D 96049 Bamberg, Germany}
\affil{$^2$Department of Astrophysics, Faculty of Science, Radboud University Nijmegen, P.O. Box 9010, 6500 GL Nijmegen, NE}
\affil{$^3$Department of Physics, University of the Western Cape, Private Bag X17, Bellville 7535, South Africa}
\affil{$^4$South African Astronomical Observatory, PO Box 9, 7935 Observatory, Cape Town, South Africa}
\affil{$^5$Southern African Large Telescope Foundation, PO Box 9, 7935 Observatory, Cape Town, South Africa}
\affil{$^6$Department of Physics, University of Warwick, Conventry CV4 7AL, UK}}

\begin{abstract}
Helium abundances and atmospheric parameters have been determined from high resolution spectra for a new sample of 46 bright hot subdwarf B (sdB) stars. The helium abundances have been measured with high accuracy. We confirm the correlation of helium abundance with temperature and the existence of two distinct sequences in helium abundance found previously. We focused on isotopic shifts of helium lines and found $^{3}$He to be strongly enriched in 8 of our programme stars. Most of these stars cluster in a small temperature range between $27\,000\,{\rm K}$ and $31\,000\,{\rm K}$ very similar to the known $^{3}$He-rich main sequence B stars, which cluster at somewhat lower temperatures. This phenomenon is most probably related to diffusion processes in the atmosphere, but poses a challenge to diffusion models.
\end{abstract}

\section{Introduction}

The atmospheric helium abundances of sdBs are poorly understood. From the typical abundance patterns of these stars, which show a depletion of light elements as well as an enrichment of heavy metals, it has been concluded that diffusion play an important role in their atmospheres. However, diffusion models predict an almost total depletion of helium in contrast to what is observed. The helium abundances of sdBs range from slightly above solar down to $\log{y}<-4$. Mass loss caused by stellar winds as well as extra mixing in the atmosphere have been invoked to counteract gravitational settling and explain the observed helium abundances \citep[see ][ and references therein]{hu11}. 

\citet{edelmann03} found a correlation of helium abundance with temperature. The hotter the sdB, the more helium is present in its atmosphere. Similar correlations have been found by other groups (see e.g. Vennes et al. these proceedings). However, \citet{edelmann03} also reported the discovery of two distinct sequences showing a similar correlation with temperature, the ''lower sequence'' being offset by about $2\,{\rm dex}$ from the upper sequence. The majority of stars lie on the ''upper sequence''. Those sequences could not be clearly identified in other datasets so far \citep[e.g. ][]{lisker05, geier11}. \citet{otoole08} combined the then published data sets and found the stars of the upper sequence to lie near the Extreme Horizontal Branch (EHB) band in the $T_{\rm eff}-\log{g}$-plane, as expected, whereas the lower-sequence stars lie in a much more dispersed area \citep[see Figs.~2,3 in][]{otoole08}.

Gravitational settling can also lead to isotopic anomalies in stellar atmospheres. In the case of helium the light isotope $^{3}$He can be enriched with respect to the usually much more abundant $^{4}$He. Such an enrichment has initially been found in main sequence B stars with subsolar helium abundance \citep{hartoog79}. However, \citet{heber91} detected strong line shifts in the sdB star SB\,290 and the blue horizontal branch star PHL\,25 indicating that basically the whole helium content of the atmosphere consists of $^{3}$He. Later on \citet{edelmann01} and \citet{heber04} found another three sdBs, where $^{3}$He is enriched in the atmosphere.

Here we present the results of a quantitative spectral analysis of a sample of 46 sdB stars from high resolution spectra.

\begin{figure}
\begin{center}
\includegraphics[width=10cm]{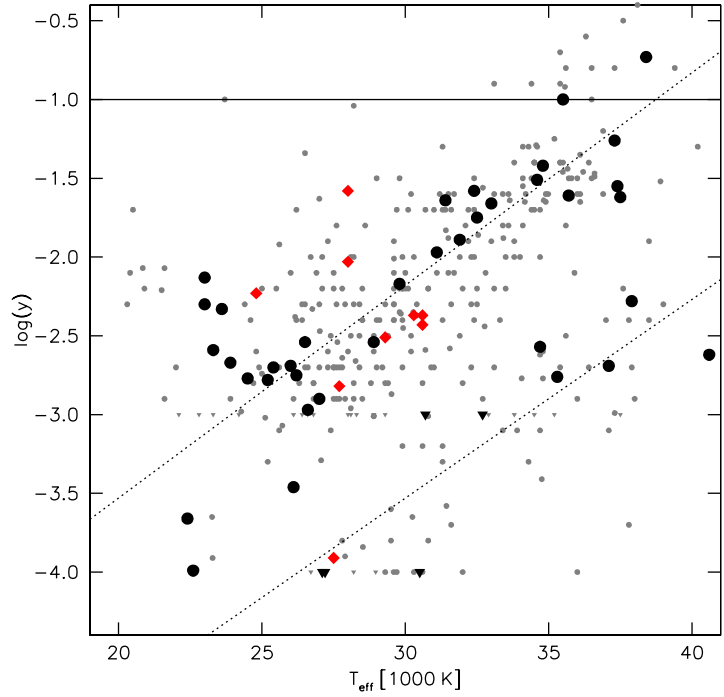}
\caption{Helium abundance $\log{y}$ plotted against effective temperature. The filled symbols mark the results from our study. Filled red diamonds mark objects where isotopic shifts due to an enrichment of $^{3}$He were detected, filled circles objects with atmospheres dominated by $^{4}$He. Upper limits are marked with triangles. The solid horizontal line is drawn at solar helium abundance. The two dotted lines are regression lines for the two distinct helium sequences taken from \citet{edelmann03}. Results taken from the literature are plotted as grey symbols \citep{saffer94, maxted01, edelmann03, morales03, lisker05, geier11, vennes11, oestensen10}.}
\label{fig:abun}
\end{center}
\end{figure}

\section{Data Analysis}

46 bright subdwarf B stars were observed with the FEROS spectrograph ($R=48\,000$, $3750-9200\,{\rm \AA}$) mounted at the ESO/MPG 2.2m telescope in La Silla. Five stars were observed with the FOCES spectrograph ($R=30\,000$, $3800-7000\,{\rm \AA}$) mounted at the CAHA 2.2m telescope. The data were reduced with the MIDAS package. Medium resolution spectra of 13 stars were obtained with the ISIS spectrograph ($R\simeq4000,\lambda=3440-5270\,{\rm \AA}$) mounted at the WHT. 13 sdBs discovered in the course of the Edinburgh-Cape blue object survey \citep{stobie97} have been observed with the grating spectrograph and intensified reticon photon counting system on the 1.9m telescope of the SAAO ($R\simeq1300,\lambda=3300-5600\,{\rm \AA}$). Spectra of five sdBs have been taken with the CAFOS spectrograph mounted at the CAHA 2.2m telescope ($R\simeq1000,\lambda=3500-5800\,{\rm \AA}$).
 
Atmospheric parameters and helium abundances have been determined by fitting model spectra to the hydrogen Balmer and helium lines \citep{heber00} of the high-resolution spectra using the SPAS routine developed by H. Hirsch. The parameter determination from the high-resolution spectra needs to be checked and systematic effects have to be quantified properly. In order to do this we analysed medium-resolution spectra in the same way as the high-resolution data and derived systematic uncertainties by comparing the results.

\begin{figure}
\begin{center}
\includegraphics[width=10cm]{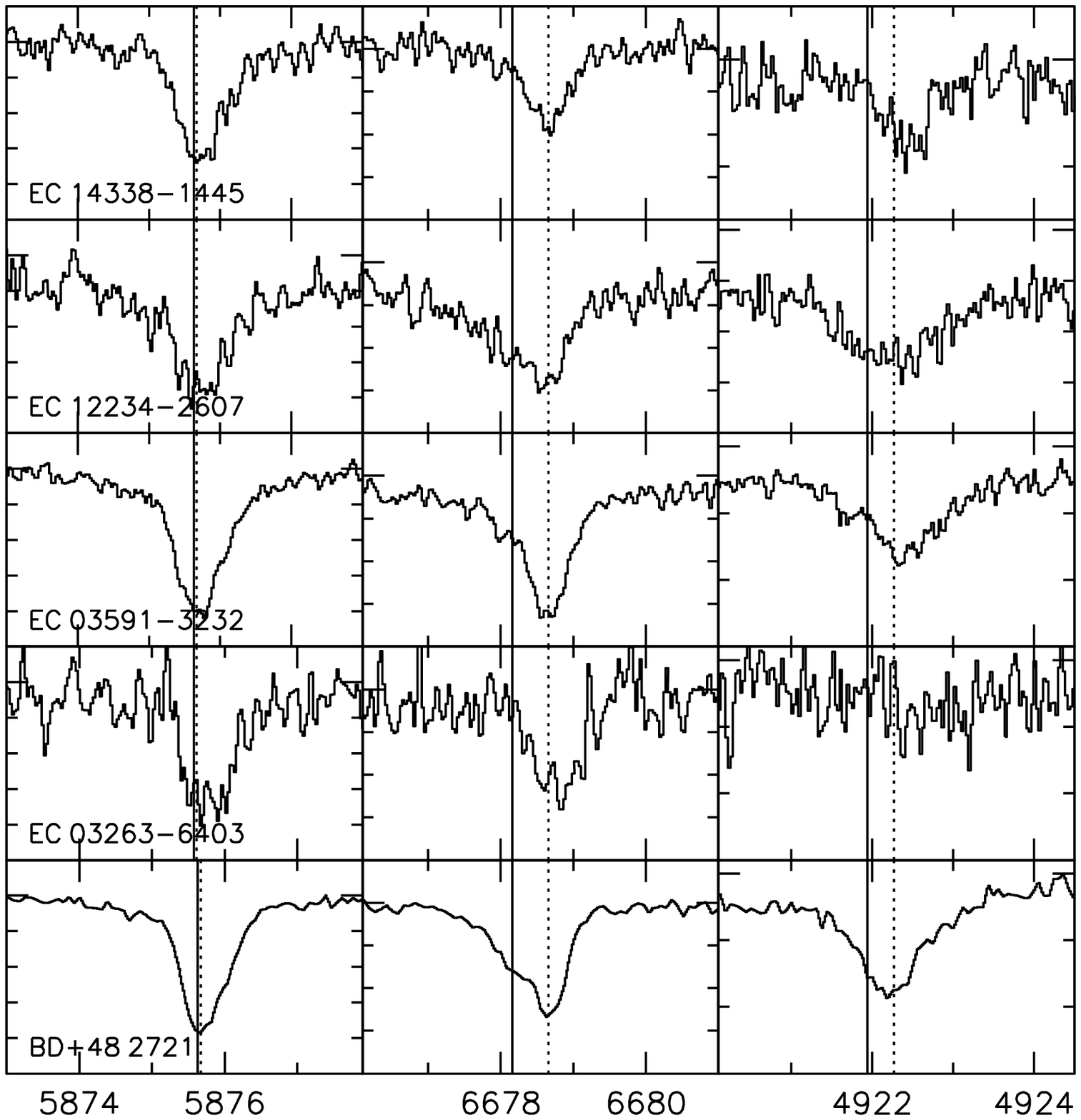}
\includegraphics[width=10cm]{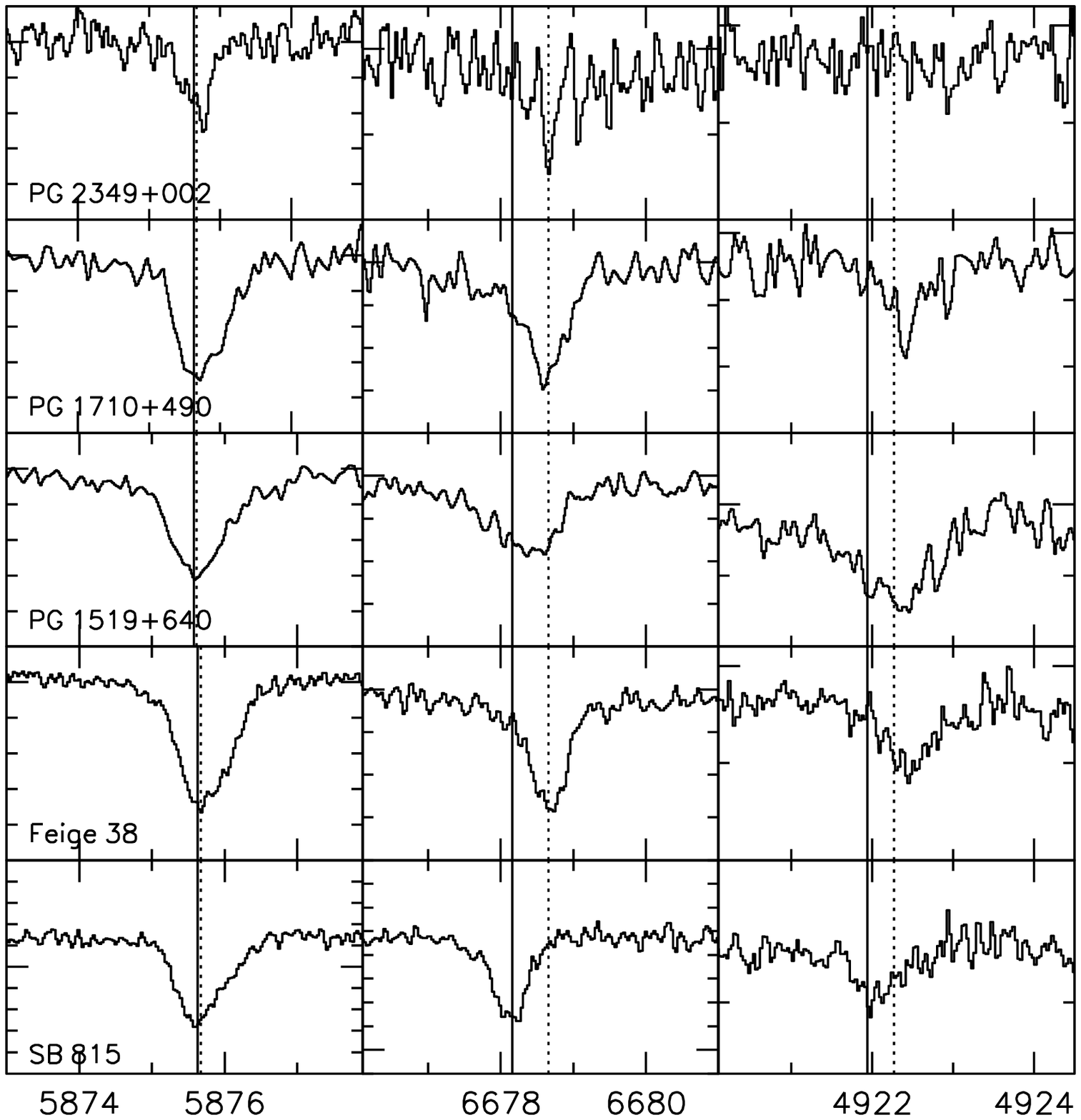}
\caption{Helium lines of sdB stars. The rest wavelengths of the $^{4}$He (solid) and $^{3}$He lines (dotted) are plotted as vertical lines.}
\label{fig:isotop}
\end{center}
\end{figure}

\section{Helium Abundances}

Fig.~\ref{fig:abun} shows the helium abundances of our sample plotted against the effective temperature. All but two of our programme stars have subsolar helium abundances typical for sdB stars. The correlation of helium abundance with temperature discovered by \citet{edelmann03} can be clearly seen as well as the two distinct sequences showing a similar correlation with temperature. 

Combining these data with the results of other studies the underlying pattern becomes apparent. In Fig.~1 our results are overplotted with the two regression lines calculated by \citet{edelmann03} and based on their results. The two lines match very well with the sequences seen in our sample. We define a dividing line between the two helium sequences at  $\log{y}=0.127\,T_{\rm eff}/1000\,{\rm K}-6.718$ Accordingly, 36 stars ($71\%$) are associated with the upper sequence while 15 ($29\%$) belong to the lower one. The respective fractions of the full sample of 349 sdBs are $77\%$ and $23\%$.

\begin{figure}
\begin{center}
\includegraphics[width=10cm]{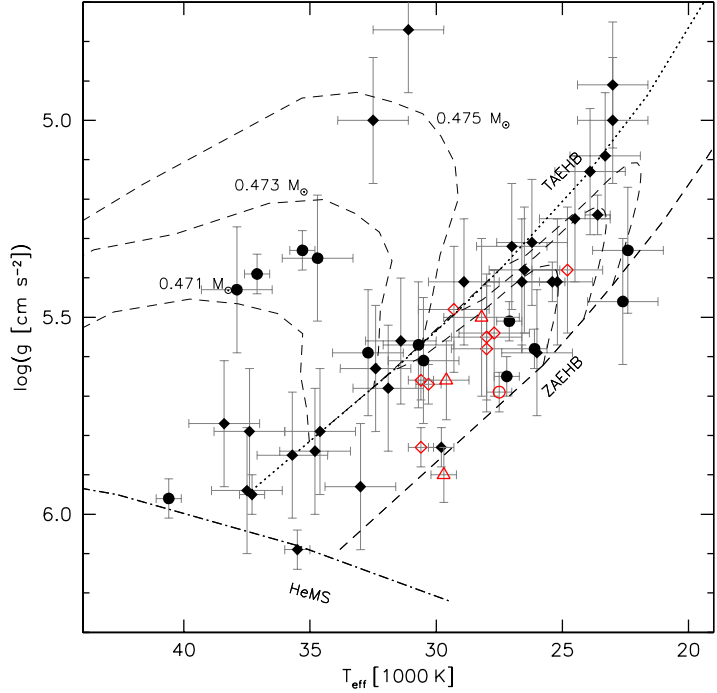}
\caption{$T_{\rm eff}-\log{g}$-diagram for the entire sample under study. The helium main sequence (HeMS) and the EHB band (limited by the zero-age EHB, ZAEHB, and the terminal-age EHB, TAEHB) are superimposed with EHB evolutionary tracks for solar metallicity taken from \citet{dorman93} labelled with their masses. Red open symbols mark objects where isotopic shifts due to an enrichment of $^{3}$He were detected, filled symbols objects with atmospheres dominated by $^{4}$He. The diamonds mark stars belonging to the upper helium sequence, the circles stars belonging to the lower sequence. The triangles mark the three sdBs with enriched $^{3}$He from the literature.}
\label{fig:tefflogg}
\end{center}
\end{figure}

\section{The $^{3}$He isotopic anomaly}

Searches for the $^{3}$He isotope in stellar atmospheres have so far been restricted to a few stars only. The high resolution spectra are perfectly suited to search for small shifts in the rest wavelengths of the helium lines due to the enrichment of $^{3}$He. Those shifts can be modelled quite accurately and show a typical pattern. While some lines like He\,{\sc i}\,5876 are only shifted by $0.04\,{\rm \AA}$ towards redder wavelengths, the shifts of He\,{\sc i}\,4922 and He\,{\sc i}\,6678 are significant \citep[$0.33$ and $0.50\,{\rm \AA}$ respectively,][]{fred51}. Displacements of this order can be easily detected in the high resolution spectra. The SPY sample \citep{lisker05} has not been studied, because the UVES spectra did not cover the sensitive He\,{\sc i}\,6678 line. 

All stars of our sample have been examined. In 8 cases isotopic shifts due to the presence of $^{3}$He are clearly visible (see Fig.~\ref{fig:isotop}). Hence $17\%$ of our programme stars show the $^{3}$He anomaly. As can be seen in Fig.~\ref{fig:abun} seven belong to the upper helium sequence, while only one star belongs to the lower sequence. This may be a selection effect, because sdBs with low helium abundance show only weak lines, which are less suited to detect isotopic shifts.

\section{Discussion}

The distribution of these stars in the $T_{\rm eff}$-$\log{g}$-diagram is shown in Fig.~\ref{fig:tefflogg} including the three sdBs with isotopic shifts taken from literature. It can be clearly seen that they cluster in a narrow temperature range between $27\,000\,{\rm K}$ and $31\,000\,{\rm K}$ with BD+48\,2721 ($T_{\rm eff}=24\,800\,{\rm K}$) being the only exception. Given the uncertainties, this $^{3}$He-strip may be pure. 

Most stars show clear shifts of the He\,{\sc i} line at $6678\,{\rm \AA}$ indicating that almost all helium in the atmosphere is $^{3}$He. BD+48\,2721, EC\,12234$-$2607 and PG\,1519+640 show strong lines of $^{3}$He blended with weak components of $^{4}$He. These three stars cover the whole $^{3}$He temperature strip. The isotope ratio is therefore not correlated to the effective temperature.

An $^{3}$He isotope anomaly has first been found for chemically peculiar main sequence stars of spectral type B. The $^{3}$He-stars were found at effective temperatures between $18\,000\,{\rm K}$ and $21\,000\,{\rm K}$ separating helium-poor stars at lower $T_{\rm eff}$ from helium-rich stars at higher $T_{\rm eff}$ \citep{hartoog79}. In Fig.~\ref{fig:tefflogg}
a similar pattern can be seen for the sdBs. The stars enriched in $^{3}$He occupy a small strip in $T_{\rm eff}$, while the helium abundance decreases towards lower temperatures and rises towards higher temperatures. \citet{michaud11} carried out diffusion calculations and predict a mild enrichment of $^{3}$He, but due to gravitational settling of the heavier isotope this should be the case in all sdBs. Hence, the $^{3}$He strip stars lacks an explanation.

\citet{hartoog79} argued that diffusion is responsible for this effect. At low temperatures the radiation pressure is not strong enough to support helium in the atmosphere. As soon as the temperature reaches a certain threshold value, the less massive $^{3}$He can be supported, but not the more abundant $^{4}$He. This leads to an enrichment of $^{3}$He in the atmosphere. At even higher temperatures both isotopes are enriched and the isotopic anomaly vanishes as the helium abundance rises.



\begin{thebibliography}{}

\bibitem[Dorman et al.(1993)]{dorman93}
Dorman, B., Rood, R. T., \& O'Connell, R. W. 1993, ApJ, 419, 596
\bibitem[Edelmann et al.(2003)]{edelmann03}
Edelmann, H., Heber, U., Hagen, H.-J., et al. 2003, A\&A, 400, 939
\bibitem[Edelmann et al.(2001)]{edelmann01}
Edelmann, H., Heber, U., \& Napiwotzki, R. 2001, AN, 322, 401
\bibitem[Fred et al.(1951)]{fred51}
Fred, M., Tomkins, F. S., Brody, J. K., \& Hamermesh, M. 1951, Phys. Rev., 82, 406
\bibitem[Geier at al.(2011)]{geier11}
Geier, S., Hirsch, H., Tillich, A., et al. 2011b, A\&A, 530, 28
\bibitem[Hartoog \& Cowley(1979)]{hartoog79}
Hartoog, M. R., \& Cowley, A. P. 1979, ApJ, 228, 229
\bibitem[Heber(1991)]{heber91}
Heber, U. 1991, IAUS, 145, 363
\bibitem[Heber(2004)]{heber04}
Heber, U., \& Edelmann, H. 2004, Ap\&SS, 291, 341
\bibitem[Heber et al.(2000)]{heber00}
Heber, U., Reid, I. N., \& Werner, K. 2000, A\&A, 363, 198
\bibitem[Hu et al.(2011)]{hu11}
Hu, H., Tout, C. A., Glebbeek, E., \& Dupret, M.-A. 2011, MNRAS, in press
\bibitem[Lisker et al.(2005)]{lisker05}
Lisker, T., Heber, U., Napiwotzki, R., Christlieb, N., Han, Z., et al. 2005, A\&A, 430, 223
\bibitem[Maxted et al.(2001)]{maxted01}
Maxted, P. F. L., Heber, U., Marsh, T. R., North, R. C., 2001, MNRAS, 326, 139 
\bibitem[Michaud et al.(2011)]{michaud11}
Michaud, G., Richer, J., \& Richard, O. 2011, A\&A, 529, 60
\bibitem[Morales et al.(2003)]{morales03}
Morales-Rueda, L., Maxted, P. F. L., Marsh, T. R., North, R. C., \& Heber, U. 2003, MNRAS, 338, 752
\bibitem[\O stensen et al.(2010)]{oestensen10}
\O stensen, R. H., Silvotti, R., Charpinet, S., et al. 2010, MNRAS, 409, 1470
\bibitem[O'Toole(2008)]{otoole08}
O'Toole, S. J. 2008, ASP Conf. Ser., 392, 67
\bibitem[Saffer et al.(1994)]{saffer94}
Saffer, R. A., Bergeron, P., Koester, D., Liebert, J. 1994, ApJ, 432, 351
\bibitem[Stobie et al.(1997)]{stobie97}
Stobie, R. S., Kilkenny, D., O'Donoghue, D., et al. 1997, MNRAS, 287, 848
\bibitem[Vennes et al.(2011)]{vennes11}
Vennes, S., Kawka, A., \& N\'emeth, P. 2011, MNRAS, 410, 2095

\end{thebibliography}

\end{document}